\documentclass[%
reprint,
superscriptaddress,
amsmath,amssymb,
aps,
prb,
floatfix,
]{revtex4-2}

\usepackage{dcolumn}

\usepackage[T1]{fontenc}
\usepackage[utf8]{inputenc}
\usepackage{bbold}
\usepackage{amsmath}
\usepackage{amssymb}
\usepackage{graphicx}
\usepackage{graphicx}
\usepackage{subscript}
\usepackage[unicode=true,pdfusetitle,
bookmarks=true,bookmarksnumbered=false,bookmarksopen=false,
breaklinks=false,pdfborder={0 0 1},backref=false,colorlinks=false]
{hyperref}
\usepackage{xcolor}
\usepackage[title]{appendix}
\usepackage{bm}
\usepackage{physics}
\usepackage[version=4]{mhchem}
\usepackage{textcomp}
\usepackage{indentfirst}
\usepackage{listings}
\usepackage{svg}
\usepackage[export]{adjustbox}
\usepackage{lipsum} 
\usepackage{braket}

\begin{document}

\title{Enabling the bulk photovoltaic effect in centrosymmetric materials through an external electric field}

\author{Guilherme J. Inacio}
\email{guijanone@gmail.com, corresponding author}
\affiliation{Departamento de Física, Universidade Federal do Espírito Santo, 29075-910 Vit\'oria-ES, Brazil}
\affiliation{Departamento de F\'isica de la Materia Condensada, Universidad Aut\'{o}noma de Madrid, E-28049 Madrid, Spain}
\author{Juan Jos\'e Esteve-Paredes}
\affiliation{Centro de F\'isica de Materiales, Universidad del Pa\'is Vasco (UPV/EHU), E-20018 Donostia-San Sebasti\'an, Spain}
\author{Maur\'icio F. C. Martins Quintela}
\affiliation{Departamento de F\'isica de la Materia Condensada, Universidad Aut\'{o}noma de Madrid, E-28049 Madrid, Spain}
\affiliation{Condensed Matter Physics Center (IFIMAC), Universidad Aut\'{o}noma de Madrid, E-28049 Madrid, Spain}
\author{Wendel S. Paz}
\affiliation{Departamento de Física, Universidade Federal do Espírito Santo, 29075-910 Vit\'oria-ES, Brazil}
\author{Juan Jos\'e Palacios}
\affiliation{Departamento de F\'isica de la Materia Condensada, Universidad Aut\'{o}noma de Madrid, E-28049 Madrid, Spain}
\affiliation{Condensed Matter Physics Center (IFIMAC), Universidad Aut\'{o}noma de Madrid, E-28049 Madrid, Spain}
\affiliation{Instituto Nicol\'as Cabrera (INC), Universidad Aut\'{o}noma de Madrid, E-28049 Madrid, Spain}

\date{\today}

\begin{abstract}
    We develop a practical approach to electrically tuning the nonlinear photoresponse of two-dimensional semiconductors by explicitly incorporating a static out-of-plane electric field into the electronic ground state prior to optical excitation, as a gate bias. The method is implemented by dressing a Wannier-interpolated Hamiltonian with the field through its position matrix elements, which allows the gate bias to modify orbital hybridization and band dispersion beyond perturbative treatments. Within the independent-particle approximation, the resulting second-order (shift) conductivity is evaluated for both centrosymmetric and non-centrosymmetric layered systems. Applied to \ce{MoS2}, the approach captures the emergence of a finite shift current in centrosymmetric bilayers and the tunability of intrinsic responses in polar structures. The shift conductivity rises linearly at small fields and saturates at higher intensities, reflecting the competition between the growing shift vector and the weakening interband coupling as resonant transitions move away from high-symmetry valleys. A Taylor expansion of the field-dressed conductivity connects this behavior to the third-order optical response, revealing a unified picture of field-induced nonlinearities. These results establish field dressing of Wannier Hamiltonians as a practical route to model and predict nonlinear photocurrents in layered materials.
\end{abstract}

\maketitle


\section{Introduction}
    The photoresponse of two-dimensional (2D) crystals has attracted sustained interest since the first demonstrations of efficient photodetection in graphene and transition-metal dichalcogenides (TMDs) \cite{wu_quantum_enhanced_2012, spanier2016power, osterhoudt2019colossal, berkelbach2013theory}. Applications span integrated photonics, photovoltaics, and ultrasensitive photodetectors \cite{sauer_shift_2023, dai_recent_2023, young2012first, chan_giant_2021}. More recently, attention has shifted toward nonlinear second-order phenomena, most notably second-harmonic generation (SHG) and the bulk photovoltaic effect (BPVE), which rely on the absence of inversion symmetry \cite{boyd2008nonlinear, puente_uriona_ab_2023, henriques_excitonic_2022, ruan_exciton_2024, taghizadeh_nonlinear_2019, quintela_tunable_2024, xiong2021atomic, wei2021electric, brun2015intense}. This has motivated different strategies to activate second-order responses in nominally centrosymmetric 2D materials, including controlled strain \cite{yu_charge-induced_2015, berciano_fast_2018, shao2019strain, mennel_second_2018}, stacking engineering and twist angles \cite{maity_atomistic_2025, gao2020tunable, shan2018stacking}, and out-of-plane electric fields \cite{zheng_gate_2023, klein2017electric}. However, while electrically tunable SHG, including electric-field-induced second-harmonic (EFISH) responses where a static field activates an effective $\chi^{(2)} \propto \chi^{(3)}E_{\mathrm{DC}}$, has been widely studied \cite{wang_giant_2024, klein2017electric, chen_gigantic_2019, brun2015intense, gu_nonlinear_2025}, the corresponding field-induced BPVE remains comparatively unexplored.
    
    A first step in this direction was taken by Zheng \textit{et al.} \cite{zheng_gate_2023}, who investigated field-induced shift and injection currents in AA$'$- and AB-stacked bilayer graphene using a four-band $\pi$-orbital tight-binding model \cite{wickramaratne_monolayer_2018, fujimoto_band_2016, brun2015intense}. Their results showed that breaking inversion symmetry via a perpendicular gate indeed generates finite dc photocurrents. Yet two limitations complicate the interpretation. First, pristine AA stacking is a semimetal, thus the applied field simultaneously open a gap and break inversion, making it difficult to separate symmetry effects from semimetal-to-semiconductor transitions. Second, the model couples the static field only to diagonal on-site terms, neglecting field-induced modifications of interlayer and extended hoppings. These off-diagonal contributions become important once the field hybridizes electronic states across layers, and their omission can mask or distort the true field-induced symmetry breaking.
    
    A complementary viewpoint was provided by Fregoso and co-workers, who developed a third-order perturbative formalism in which one of the driving fields carries zero frequency \cite{fregoso_jerk_2018, fregoso_bulk_2019}. In this approach, the dc photocurrent in a crystal under static field is encoded in the mixed third-order tensor $\sigma^{(3)}_{abcd}(0;\omega,-\omega,0)$, from which field-induced shift, injection, and jerk currents can be obtained. This framework is formally elegant and captures the leading linear-in-field response. However, because the static field enters only perturbatively, it cannot describe band-gap renormalization \cite{castro_biased_2007, zaabar_effects_2023}, field-induced hybridization of Bloch states \cite{drummond_electrically_2012, chakrabarty_fate_2025, de2024floquet}, or the eventual saturation and non-monotonic evolution of $\sigma^{(2)}$ \ at experimentally reasonable field strengths \cite{yu_highly_2013, slobodyan2022analysis, wang_electric-field_2018, kovalchuk_revealing_2025, weintrub_generating_2022, fillion2022gate, hiraoka_terahertz_2025, rashidi_tuning_2024, green20163}. Moreover, only a single field-linear value of the effective $\sigma^{(2)}$ is obtained, leaving the full evolution from the perturbative to the strong-field regime inaccessible.
    
    In this work, we demonstrate that a gate bias, applied as a static out-of-plane electric field, can induce and continuously tune shift-current responses in layered materials, including centrosymmetric bilayer crystals such as 2H \ce{MoS2}. Our approach incorporates the field directly through matrix elements of the position operator in a Wannier-interpolated Hamiltonian. Here, intersite (off-diagonal) dipole matrix elements arise naturally \cite{resta1998quantum, esteve_paredes_comprehensive_2023, mostofi2014updated, ozaki2024closest}, allowing the field to modify not only the local potential but also orbital hybridization and electronic dispersion. This enables a fully non-perturbative treatment in the static field, capturing the onset, evolution, and eventual saturation of the shift conductivity. The method applies equally to centrosymmetric (2H) and non-centrosymmetric (3R) bilayers, as well as to monolayers with finite thickness along $z$.
    By expanding $\sigma^{(2)}(\omega;E_{\mathrm{DC}}^{z})$ in powers of the field, we recover analytically the mixed third-order tensor $\sigma^{(3)}_{abcd}(0;\omega,-\omega,0)$ obtained in Fregoso \emph{et al.}'s work \cite{fregoso_jerk_2018,fregoso_bulk_2019}, establishing a direct connection between our non-perturbative field dressing and the perturbative third-order formalism. This provides a unified description of weak- and strong-field regimes and clarifies the origin of saturation at large field.
    
    Section \ref{sec:theory} introduces the formalism, including the implementation of the static field in the Wannier basis, highlighting the important role of off-diagonal position-matrix elements not present in minimal tight-binding models. It also treats the evaluation of the shift current, and its connection to third-order response theory. Section \ref{sec:MoS2} focuses on \ce{MoS2}, where we analyze the monolayer, the centrosymmetric AA$'$ (2H) bilayer, and the non-centrosymmetric AB (3R) bilayer, showing how the magnitude and tensor structure of the second-order response evolve under static field. Computational parameters and Wannierization details are provided in Appendix \ref{sec:methods}.

\section{Theory\label{sec:theory}}

\subsection{Hamiltonian with static field}
    Following the length-gauge formulation of Aversa-Sipe and Sipe-Shkrebtii \cite{aversa_nonlinear_1995,sipe2000second}, previously applied in related contexts \cite{esteve-paredes_excitons_2025,uria2023efficient}, we consider a two-dimensional crystal whose principal axes lie in the $xy$ plane. A static out-of-plane field $\mathbf{E}_{\mathrm{DC}}=(0,0,E_{z})$ breaks inversion symmetry, while an in-plane monochromatic probe field $E_a^{(\mathrm{ac})}e^{i\omega t}$ ($a=x,y$) excites the system. In the length gauge, the total Hamiltonian reads
    \begin{align}\label{eq:field-dressing}
        \hat H
        = \hat H_0
          - eE_{\mathrm{DC}}\,\hat r_z
          - eE_a^{(\mathrm{ac})}e^{i\omega t}\,\hat r_a,
    \end{align}
    where $\hat H_0$ is the mean-field equilibrium Hamiltonian and $\hat{\mathbf r}$ is the position operator. To treat the static field non-perturbatively, we define
    \begin{align}
        &\hat H_0^{(\mathrm{DC})}
          = \hat H_0 - eE_{\mathrm{DC}}\,\hat r_z, \\
        &\hat H'
          = - eE_a^{(\mathrm{ac})}e^{i\omega t}\,\hat r_a,
    \end{align}
    and regard $\hat H'$ as a first-order optical perturbation.
    
    It is instructive to inspect the matrix elements of the position operator in the Bloch basis $\{\ket{m\mathbf{k}}\}$ of $\hat H_0$,
    \begin{align}
        r^{z}_{mn\mathbf{k}}
        = \braket{m\mathbf{k}|\hat r_z|n\mathbf{k}}.
    \end{align}
    Because $z$ is non-periodic, these matrix elements are gauge independent and free of the Berry-phase ambiguity that affects in-plane components \cite{resta1998quantum}. Using a localized-orbital representation
    $\ket{\alpha\mathbf{k}}=\sum_{\mathbf{R}}e^{i\mathbf{k}\cdot\mathbf{R}}\ket{\alpha\mathbf{R}}$, with orbital index $\alpha$ and lattice vector $\mathbf{R}$, the position matrix elements can be decomposed into on-site and intersite contributions. For the $z$ component one can write
    \begin{align}\label{eq:onsite_intersite_decomp}
        r^{z}_{mn\mathbf{k}}
        &= r^{z,\mathrm{on}}_{nn\mathbf{k}}
         + r^{z,\mathrm{inter}}_{mn\mathbf{k}},
    \end{align}
    where the on-site part depends only on the $z$-coordinates of the orbital centers $d_z^{(\alpha)}$,
    \begin{align}
        r^{z,\mathrm{on}}_{mn\mathbf{k}}
        = \sum_{\alpha}
          c^{(m)\ast}_{\alpha\mathbf{k}} c^{(n)}_{\alpha\mathbf{k}}\,
          d_z^{(\alpha)},
    \end{align}
    and $r^{z,\mathrm{inter}}_{mn\mathbf{k}}$ collects all remaining intersite terms involving overlaps
    $\braket{\alpha\mathbf{0}|\hat r_z|\alpha'\mathbf{R}}$ with either $\mathbf{R}\neq\mathbf{0}$ or $\alpha\neq\alpha'$. In minimal tight-binding models built from point-like planar orbitals, $r^{z,\mathrm{inter}}_{mn\mathbf{k}}$ is often neglected and $d_z^{(\alpha)}$ may be identical for all orbitals in a given layer, so the static field effectively enters only as a layer-resolved on-site potential. In contrast, Wannier Hamiltonians constructed from \textit{ab initio} calculations naturally generate finite intersite dipole matrix elements \cite{resta1998quantum,esteve_paredes_comprehensive_2023,ozaki2024closest,oiwa_symmetry-adapted_2025}, allowing the static field to modify both orbital hybridization and band dispersion. This distinction is crucial in planar systems or in centrosymmetric multilayers, where purely on-site models can artificially suppress the coupling to the static field.
    
\subsection{Second-order dc response}
    The dc photocurrent to second-order in the optical field can be written as
    \begin{align}
        j_a(0)
        = \sum_{bc}\sigma^{(2)}_{abc}(\omega;E_{\mathrm{DC}})\,
          E_b^{(\mathrm{ac})}E_c^{(\mathrm{ac})},
    \end{align}
    where $\sigma^{(2)}_{abc}(\omega;0)$ reduces to the usual second-order conductivity containing shift and injection contributions \cite{sipe2000second}. In our approach, the dependence on the static field is fully contained in the eigenvalues and eigenstates of the field-dressed Hamiltonian,
    \begin{align}\label{eq:eig_DC}
        \hat H_0^{(\mathrm{DC})}\ket{n\mathbf{k}}
        = \varepsilon_{n\mathbf{k}}\ket{n\mathbf{k}}.
    \end{align}
    We obtain these eigenstates by expanding either in the Bloch basis $\{\ket{m\mathbf{k}}\}$ or, equivalently, in the Wannier basis $\{\ket{\alpha\mathbf{k}}\}$. In the Bloch representation, Eq. \eqref{eq:eig_DC} becomes
    \begin{align}
        \sum_{n}
        \Bigl[\varepsilon^{(0)}_{n\mathbf{k}}\delta_{mn}
          - eE_{\mathrm{DC}}\, r^{z}_{mn\mathbf{k}}\Bigr]
          c^{(n)}_{n\mathbf{k}}
        = \varepsilon_{n\mathbf{k}} c^{(n)}_{m\mathbf{k}},
    \end{align}
    with $r^{z}_{mn\mathbf{k}}$ as in Eq. \eqref{eq:onsite_intersite_decomp}. The resulting eigenvalues $\varepsilon_{n\mathbf{k}}$ and eigenvectors $\ket{n\mathbf{k}}$ define a new equilibrium ground state in the presence of the static field, on top of which the optical perturbation is treated to first order.
    
    Using the field-dressed eigenstates in the length-gauge formalism of Sipe and Shkrebtii \cite{sipe2000second}, the shift contribution to the BPVE can be expressed as
    \begin{align}\label{eq:shiftcurrent_shiftvector}
        &\sigma_{abc}^{(\mathrm{shift})}(\omega;E_{\mathrm{DC}})
        = \\
        &\;\;-\frac{\pi e^3}{\hbar^2}
          \sum_{mn}\!f_{mn}\,
          \left( R^{c a}_{m n} - R^{c b}_{n m} \right)
            \, r^{b}_{n m}\, r^{a}_{m n}
          \delta(\omega_{nm}-\omega),\notag
    \end{align}
    where $f_{mn}=f_{m}-f_{n}$ is the occupation difference, $r_{mn}^{b}$ are dipole matrix elements in the field-dressed basis, $\omega_{nm}=(\varepsilon_{n}-\varepsilon_{m})/\hbar$, and $R_{nm}^{a,b}$ is the shift vector. All these quantities depend implicitly on $E_{\mathrm{DC}}$ through Eq. \eqref{eq:eig_DC}. For time-reversal-symmetric systems and linearly polarized light, the injection contribution vanishes and the real part of $\sigma^{(2)}$ coincides with the shift conductivity \cite{sipe2000second}; this is the quantity we focus on throughout.
    
\subsection{Connection with third-order response}
    
    For our purposes it is sufficient to retain only those contributions to the dc current that depend on the optical field. In frequency space, one can write schematically
    \begin{align}\label{eq:J_full}
        J^a_{\mathrm{DC}}
        &= \sigma^{(2)}_{abc}(0;\omega,-\omega)\,
           E_b(\omega)E_c(-\omega) \notag\\
        &\quad + \sigma^{(3)}_{abcd}(0;\omega,-\omega,0)\,
           E_b(\omega)E_c(-\omega)E_d(0)\notag\\
        &\quad+ \ldots,
    \end{align}
    where $E_d(0)$ denotes the static field component and $\sigma^{(3)}_{abcd}(0;\omega,-\omega,0)$ encodes the static field-induced shift, injection, and jerk currents \cite{fregoso_jerk_2018,fregoso_bulk_2019}. In our scheme, the second-order photoconductivity $\sigma^{(2)}_{abc}(\omega;E_{\mathrm{DC}})$ is computed non-perturbatively with respect to $E_{\mathrm{DC}}$ via the field-dressed Hamiltonian. Assuming that $\sigma^{(2)}_{abc}(\omega;E_{\mathrm{DC}})$ is smooth around $E_{\mathrm{DC}}=0$, we can write a Taylor expansion
    \begin{align}\label{eq:taylor}
        \sigma^{(2)}_{abc}(\omega;E_{\mathrm{DC}})
        &= \sigma^{(2,0)}_{abc}(\omega)
         + E_{\mathrm{DC}}\,\sigma^{(2,1)}_{abc}(\omega)
         + \mathcal{O}\!\left(E_{\mathrm{DC}}^{2}\right),
    \end{align}
    with
    \begin{align}
        \sigma^{(2,1)}_{abc}(\omega)
        = \left.
          \frac{\partial}
               {\partial E_{\mathrm{DC}}}
          \sigma^{(2)}_{abc}(\omega;E_{\mathrm{DC}})
          \right|_{E_{\mathrm{DC}}=0}.
    \end{align}
    Since in our setup the static field points along $z$, this first-order coefficient is directly related to the mixed third-order tensor via
    \begin{align}
        \sigma^{(3)}_{abcz}(0;\omega,-\omega,0)
        = \sigma^{(2,1)}_{abc}(\omega),
    \end{align}
    so that the linear-in-field part of our non-perturbative $\sigma^{(2)}_{abc}(\omega;E_{\mathrm{DC}})$ reproduces the components $\sigma^{(3)}_{abcz}$ of the fourth-rank tensor derived in Refs. \cite{fregoso_jerk_2018,fregoso_bulk_2019}. Because the static field is applied strictly along $z$, our calculations probe this particular tensor slice, with $a,b,c\in\{x,y,z\}$ and fixed static field index $z$.
    
    In centrosymmetric crystals, $\sigma^{(2,0)}_{abc}(\omega)=0$ by symmetry, so the leading nonzero shift current at small $E_{\mathrm{DC}}$ is linear in the static field and entirely governed by $\sigma^{(3)}_{abcz}(0;\omega,-\omega,0)$. This behavior will become clear in our results below, where $\sigma^{(2)}_{abc}(\omega;E_{\mathrm{DC}})$ is strictly zero at $E_{\mathrm{DC}}=0$ and grows linearly at weak fields. At larger $E_{\mathrm{DC}}$, deviations from linearity signal the onset of higher-order terms in Eq. \eqref{eq:taylor}, which are fully captured by the field-dressed Hamiltonian and are responsible for the saturation and eventual suppression of the shift current observed in our numerical results.

\subsection{Symmetry Analysis}

    Before presenting the numerical results it is useful to summarize how a perpendicular static field modifies the point-group symmetry of the systems considered. The symmetry fixes which components of the second-order conductivity tensor $\sigma^{(2)}_{abc}$ \cite{garcia-blazquez_shift_2023} can be nonzero and explains the different qualitative responses of monolayer, 2H bilayer, and 3R bilayer \ce{MoS2}.
    
    \subsubsection{Monolayer}
    At zero static field the monolayer belongs to the $D_{3h}$ point group. The horizontal mirror $\sigma_h$ forbids all tensor components that contain an odd number of $z$ indices, so the allowed shift conductivity is restricted to in-plane components. A perpendicular static field $E_{\mathrm{DC}}$ removes $\sigma_h$ and all operations that reverse the $z$ axis, such as the $C_2'$ rotations and the $S_3$ rotoinversions. The remaining operations are the threefold rotation $C_3$ and the three vertical mirrors $\sigma_v$, which form the $C_{3v}$ point group, with non-zero components arranged as in Eq. \ref{eq:tensor_components}. Under this reduced symmetry all components permitted by $C_{3v}$ are allowed, including those that involve $z$\cite{garcia-blazquez_shift_2023}:

    \begin{flalign}\label{eq:tensor_components}
        &\sigma^{x;xx} = -\sigma^{x;yy} = -\sigma^{y;yx}, \\
        &\sigma^{z;zz},  \notag\\
        &\sigma^{x;xz} = \sigma^{y;yz}, \notag\\
        &\sigma^{z;xx} = \sigma^{z;yy}.  \notag
    \end{flalign}
    In practice, the new components remain small because the monolayer is thin along $z$, and the field can only induce a limited out-of-plane charge displacement.
     
    \subsubsection{2H bilayer (AA$'$ stacking)}
    The 2H bilayer belongs to the centrosymmetric $D_{3d}$ point group at zero field. Inversion symmetry enforces $\sigma^{(2)}_{abc}=0$ for all tensor components in equilibrium. When a perpendicular static field is applied, inversion and the rotoinversion operations are broken, while the threefold axis and the three vertical mirrors are preserved. The symmetry is therefore reduced from $D_{3d}$ to $C_{3v}$, which is the same point group as the field-dressed monolayer. In this situation all $C_{3v}$-allowed tensor components become finite. The difference with the monolayer is not the existence of additional symmetry channels but the way the field polarizes the system. In the bilayer, the two layers are related by inversion at zero field and are separated along $z$. Once the inversion is broken, the field induces opposite charge displacements in the two layers and generates a sizable shift conductivity, both in in-plane components and in mixed components that involve $z$. This explains why the field-induced response of the 2H bilayer is much stronger than the small corrections that appear in the monolayer.
    
    \subsubsection{3R bilayer (AB stacking)}
    The 3R bilayer lacks inversion and also lacks the horizontal mirror already at zero static field, so it belongs to the $C_{3v}$ point group in equilibrium. As a result, the shift current is nonzero even without external static field and the nonvanishing components follow the $C_{3v}$ pattern. A perpendicular static field keeps the point group unchanged, so the set of symmetry-allowed tensor components is the same. What changes is the built-in layer asymmetry. The stacking already produces an intrinsic potential difference between the two layers. Depending on its sign, the external field can either reinforce the built-in polarity or counteract it. These competing contributions allow the shift conductivity to be tuned in opposite directions, and in some cases the effective asymmetry can be significantly reduced. The numerical results in Sec. \ref{sec:MoS2} show this behavior clearly in components such as $\sigma^{(2)}_{zzz}$. The allowed tensor components take the form 
    
    Overall, a perpendicular field brings all three systems into a common $C_{3v}$ framework, each via a distinct symmetry-breaking route. The monolayer starts from $D_{3h}$ and acquires $C_{3v}$ under a static field, which activates additional components that remain relatively small. The 2H bilayer starts from $D_{3d}$, has no second-order response in the absence of a field, and develops a strong shift conductivity once inversion is broken. The 3R bilayer is already $C_{3v}$ at zero field, and the external static field tunes an intrinsic shift current by either amplifying or compensating the built-in layer polarity. This unified symmetry picture underlies the distinct field dependences observed in the numerical results.

\section{Results}\label{sec:MoS2}

\subsection{\ce{MoS2} Monolayer}

    As discussed in the symmetry analysis, the monolayer acquires $C_{3v}$ symmetry under a perpendicular static field, which activates tensor components containing out-of-plane indices. However, the effect is expected to be weak because the monolayer is thin along $z$ and the induced charge redistribution is limited. Here, and throughout the cases presented in this section, our analysis therefore focuses on photon energies near the bandgap, where the response is well converged and of primary physical interest, while higher-energy features may be affected by the finite $k$ mesh.
    
    \begin{figure}[ht]
        \centering
        \includegraphics[width=1\linewidth]{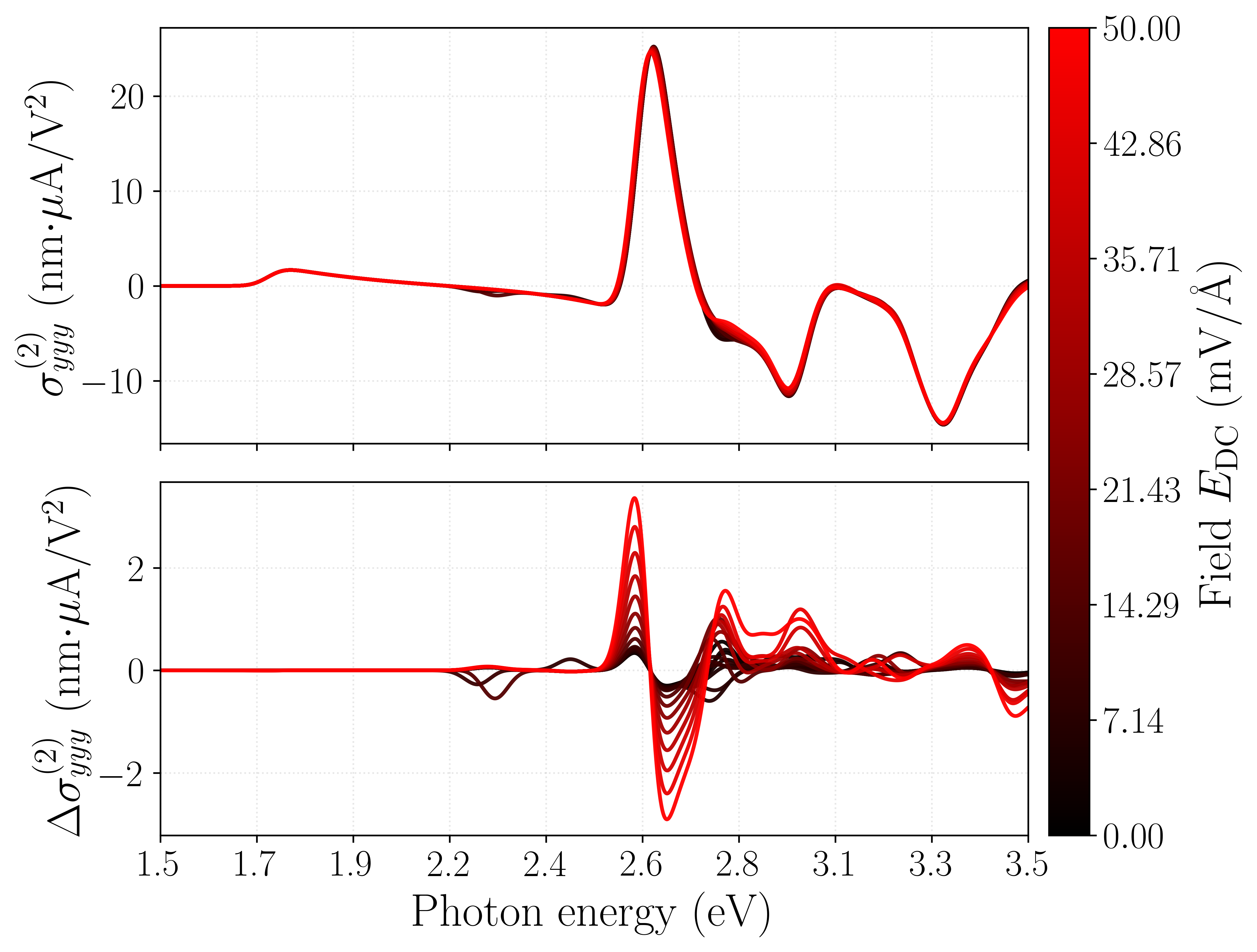}
        \caption{
            second-order conductivity $\sigma^{(2)}_{yyy}(\omega;E_{\mathrm{DC}})$ of monolayer \ce{MoS2} under static out-of-plane fields up to $0.050$ mV/\AA. 
            The top panel shows the full spectrum and the bottom panel shows the differential response $\Delta\sigma^{(2)}_{yyy}(\omega;E_{\mathrm{DC}})$. 
            Field-induced changes remain weak, on the order of $2.5$ nm$\cdot\mu$A/V$^2$.
        }
        \label{fig:mos2_mono_yyy}
    \end{figure}
    
    To quantify the effect of the static field we examine the differential response $\Delta\sigma(\omega;E_{\mathrm{DC}})=\sigma(\omega;E_{\mathrm{DC}})-\sigma(\omega;0)$. In-plane components such as $\sigma^{(2)}_{yyy}$ are already allowed at zero field and remain essentially unchanged, as shown in Fig. \ref{fig:mos2_mono_yyy}. This behavior agrees with SHG measurements where the monolayer signal varies weakly under gating while bilayers show a strong dependence \cite{klein2017electric}. This is also consistent with DFT studies showing that the monolayer bandgap is nearly insensitive to a perpendicular static field \cite{liu_tuning_2012}.
    
    \begin{figure}[ht]
        \centering
        \includegraphics[width=0.9\linewidth]{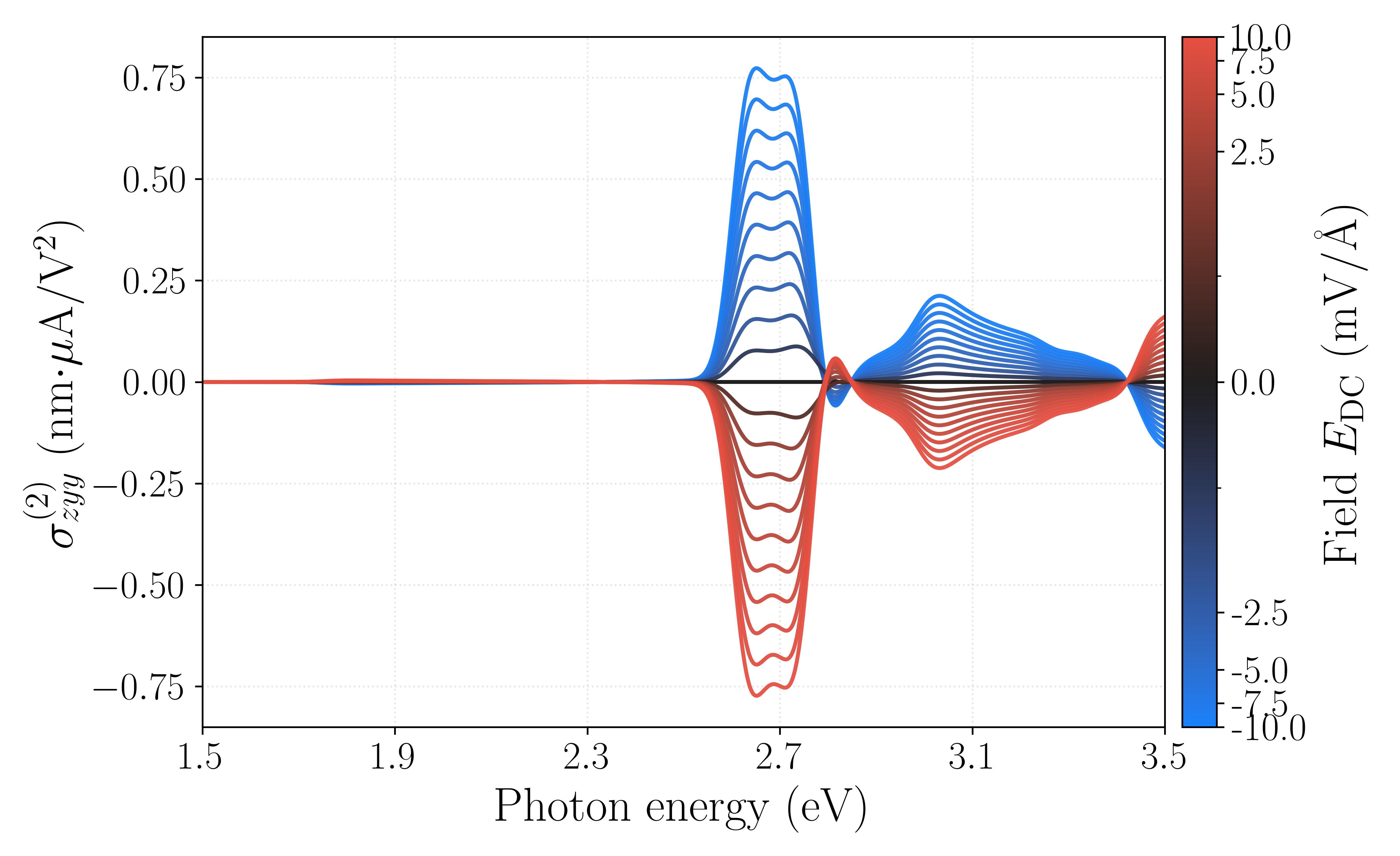}
        \caption{
            Representative out-of-plane component of the shift conductivity tensor in monolayer \ce{MoS2} under a perpendicular static field.
        }
        \label{fig:mos2_monolayer_zyy}
    \end{figure}
    
    The situation changes for tensor elements involving out-of-plane indices. As shown in Fig. \ref{fig:mos2_monolayer_zyy}, the perpendicular field activates components such as $\sigma^{(2)}_{zyy}$, $\sigma^{(2)}_{yyz}$, and $\sigma^{(2)}_{zzz}$. These components display an antisymmetric dependence on $E_{\mathrm{DC}}$, consistent with the odd character of the perturbation $-eE_{\mathrm{DC}}\hat r_z$. In the small-field regime with $|E_{\mathrm{DC}}|\lesssim 5$ meV/\AA, the behavior is linear and follows the Taylor expansion discussed in Sec. \ref{sec:theory}. Beyond this regime, at larger fields with $|E_{\mathrm{DC}}|\gtrsim 50$ meV/\AA, the curves saturate, which reflects the limited out-of-plane polarizability and the nonlinear reshaping of the band structure.
    Among the newly activated components, $\sigma^{(2)}_{zxx}$ and $\sigma^{(2)}_{zyy}$ exhibit the largest variations. Their magnitudes exceed those of $\sigma^{(2)}_{zzz}$ and $\sigma^{(2)}_{yyz}$ by well over an order of magnitude, highlighting a strong coupling between in-plane optical excitation and out-of-plane charge redistribution.

\subsection{2H-\ce{MoS2} Bilayer}

    As summarized in the symmetry analysis, the 2H (AA$'$) bilayer belongs to the centrosymmetric $D_{3d}$ point group at zero field and therefore has vanishing second-order conductivity in equilibrium. A perpendicular static field $E_{\mathrm{DC}}$ breaks inversion and lowers the symmetry to $C_{3v}$, which activates all tensor components allowed by this point group. In contrast to the monolayer, where the field mainly generates small out-of-plane components, the 2H bilayer develops sizeable in-plane and mixed components once inversion is broken, leading to a strong field-induced shift current.
    
    \begin{figure}[ht]
        \centering
        \begin{minipage}{0.9\linewidth}
        \centering
        \includegraphics[width=\linewidth]{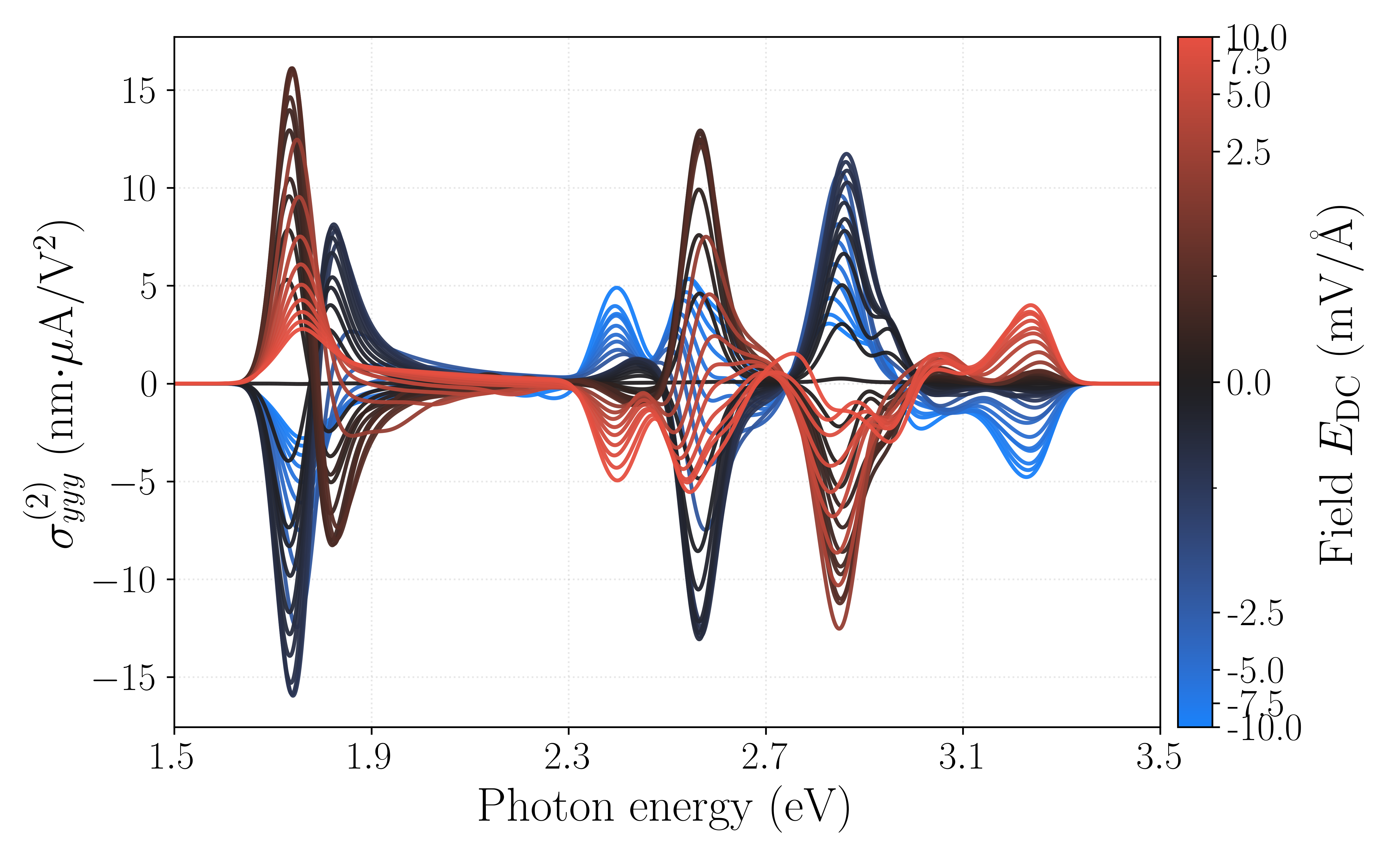}
        \end{minipage}
    
        \vspace{-1.5em}
    
        \begin{minipage}{0.9\linewidth}
            \centering
            \includegraphics[width=\linewidth]{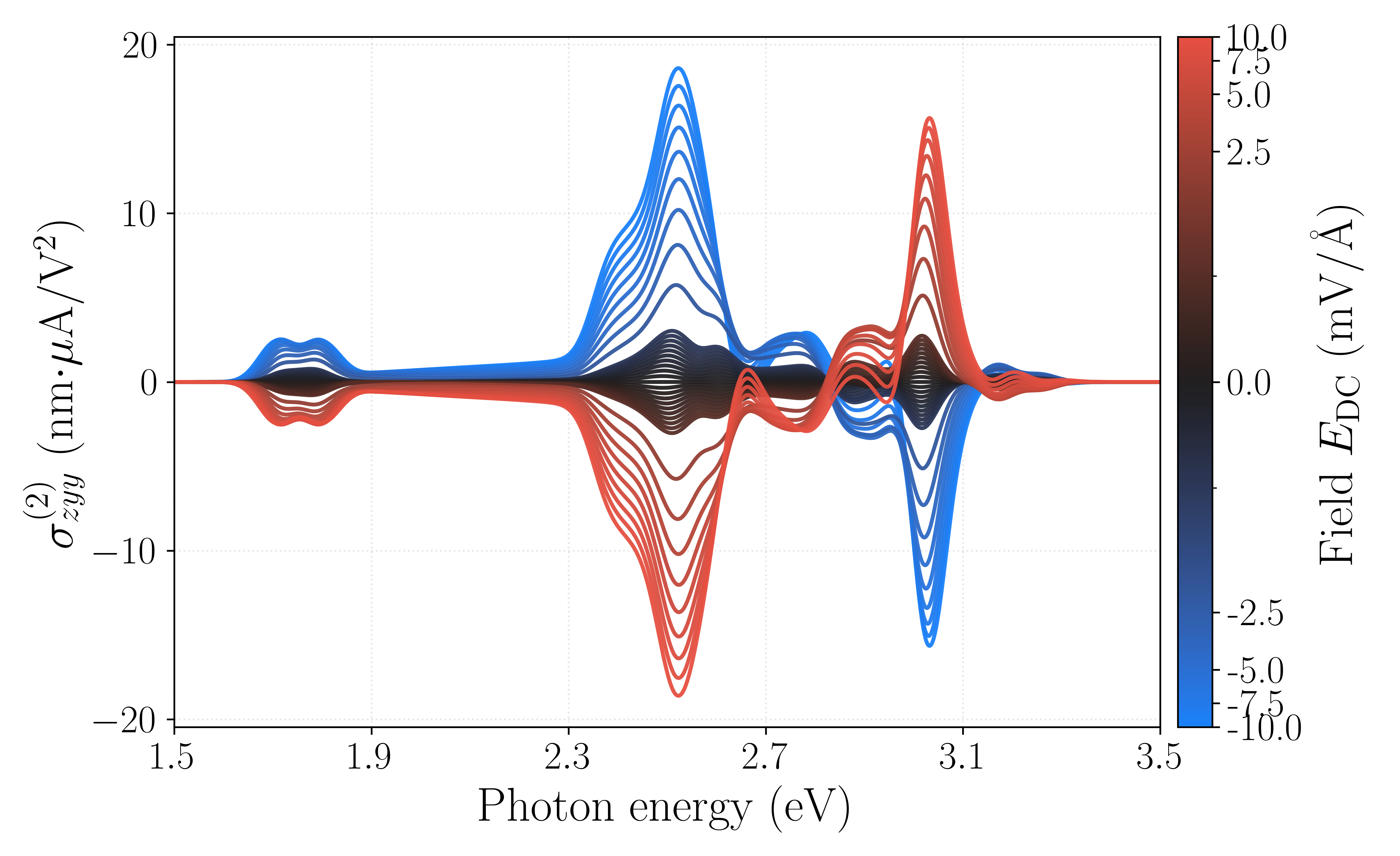}
        \end{minipage}
        \caption{
        In-plane (top) and out-of-plane (bottom) shift conductivity for 2H-\ce{MoS2} under a perpendicular static field $E_{\mathrm{DC}}$.
        }
        \label{fig:mos2_AA}
    \end{figure}
    
\begin{figure*}
    \centering

    \includegraphics[scale=0.49]{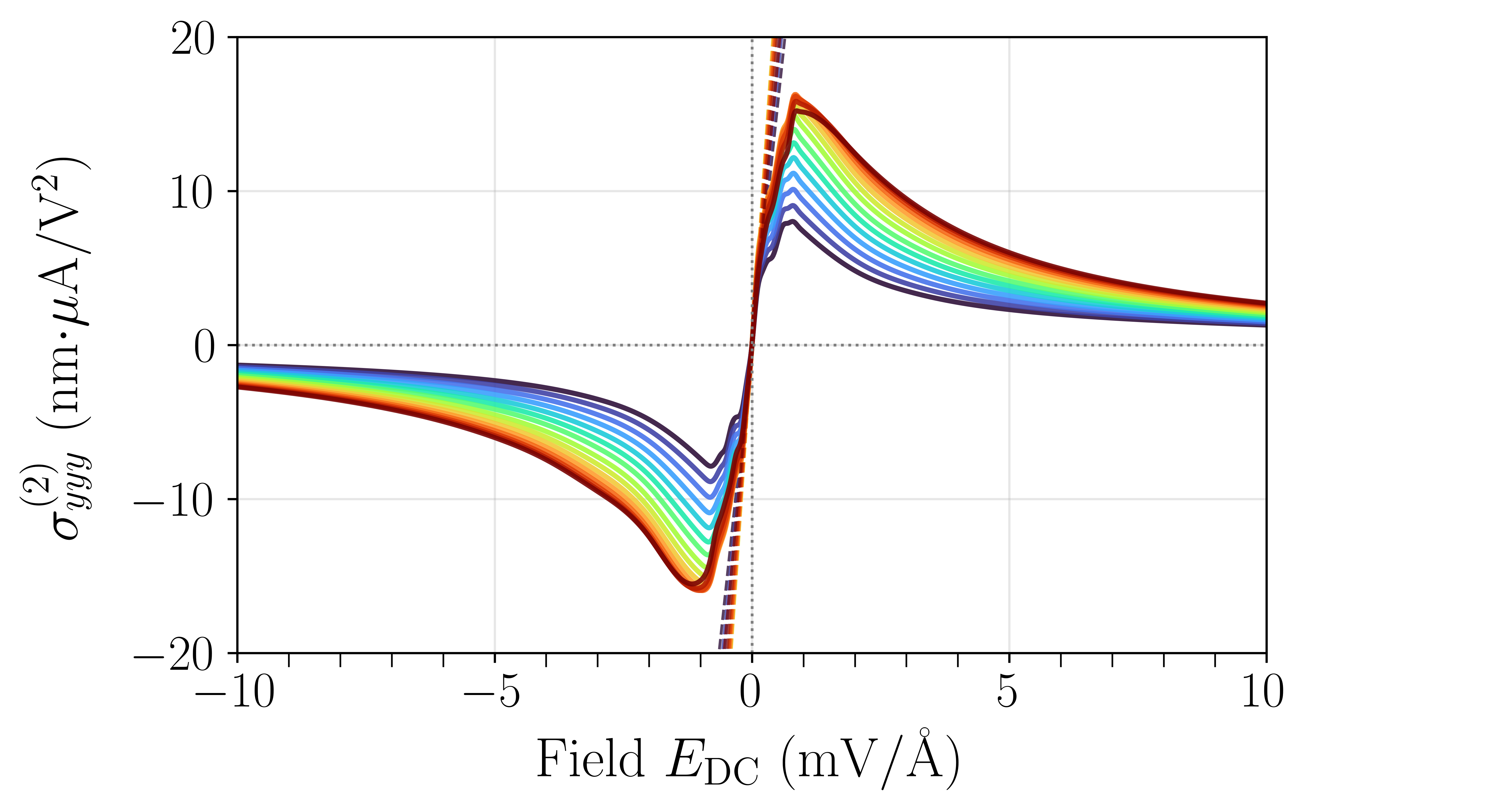}%
    \hspace{-15mm}
    \includegraphics[scale=0.49]{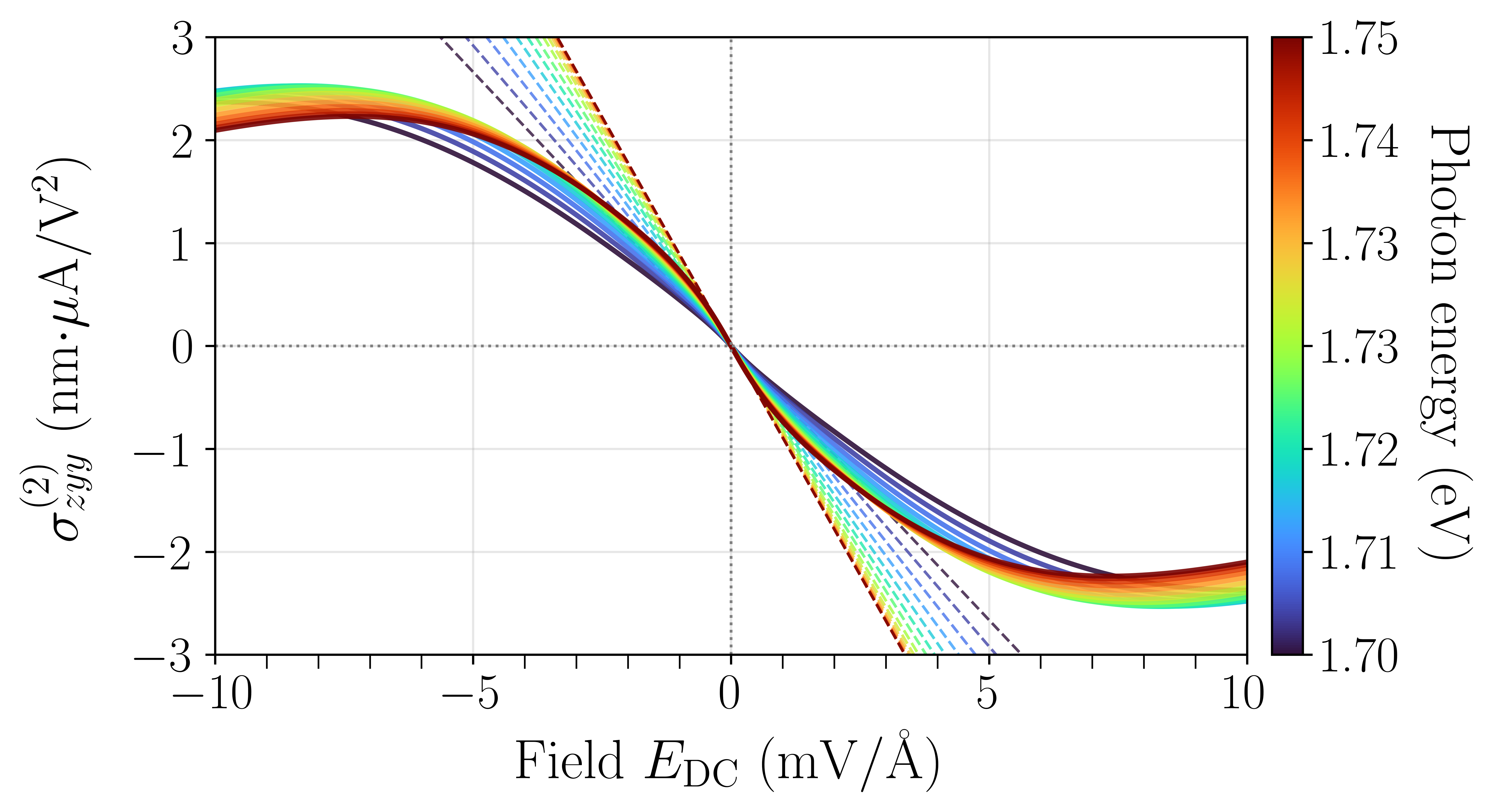}
    \caption{
        Field dependence of the second-order conductivity for 2H-\ce{MoS2}. 
        Left panels: $\sigma^{(2)}_{yyy}(\omega;E_{\mathrm{DC}})$ and its derivative with respect to $E_{\mathrm{DC}}$ for several photon energies near the band edge. 
        Right panels: corresponding results for $\sigma^{(2)}_{zyy}(\omega;E_{\mathrm{DC}})$. 
        The dashed lines show the linear behavior extracted from the derivative at $E_{\mathrm{DC}}=0$, which corresponds to the mixed third-order conductivity $\sigma^{(3)}_{abcz}(0;\omega,-\omega,0)$ through Eq.~\eqref{eq:taylor}. 
        The linear regime around zero field verifies the Taylor expansion, while deviations at larger fields reveal the onset of higher-order field dependence.
        }
    \label{fig:linearity}
    \end{figure*}

    Figure \ref{fig:mos2_AA} shows the evolution of $\sigma^{(2)}_{yyy}(\omega;E_{\mathrm{DC}})$ and $\sigma^{(2)}_{zyy}(\omega;E_{\mathrm{DC}})$ under finite field. At $E_{\mathrm{DC}}=0$ the conductivity vanishes identically for all tensor components and a finite response appears immediately once inversion is broken. The overall response increases with the static field. Among the components that involve an out-of-plane index, the mixed component $\sigma^{(2)}_{yzz}$ dominates, being significantly larger than $\sigma^{(2)}_{zzy}$ and $\sigma^{(2)}_{zzz}$. A more detailed review of the low-field regime and the  saturation onset is given in Fig.~\ref{fig:linearity}, discussed below.
    
    The origin of this saturation can be understood by examining the three ingredients that enter the shift conductivity, namely the shift vector $\mathbf{R}$, the oscillator strength $\mathcal{A}=v^a_{cv}v^b_{vc}$, and the set of resonant $\mathbf{k}$ points selected by the condition $E_{cv}(\mathbf{k})=\hbar\omega$. At small fields the shift vector grows steadily with $E_{\mathrm{DC}}$ and follows the initial increase of $\sigma^{(2)}$. At the same time the oscillator strength decreases because the field polarizes the charge density along $z$, which reduces the overlap between conduction and valence states. The band structure is also modified by the field. As the gap narrows, the resonant $\mathbf{k}$ points that satisfy $E_{cv}=\hbar\omega$ move away from the original valley and form a ring where $\mathcal{A}$ is intrinsically weaker. The apparent reduction in $\mathcal{A}$ is therefore dominated by the migration of resonant momenta rather than by a direct suppression of the velocity matrix elements. The growth of $\mathbf{R}$ and the weakening of $\mathcal{A}$ compete. At sufficiently large $E_{\mathrm{DC}}$ the latter effect dominates, so the conductivity stops increasing and eventually decreases, which marks the onset of the saturation regime.
    
    Among the newly activated components, $\sigma^{(2)}_{zxx}=\sigma^{(2)}_{zyy}$ shows the largest amplitude. Its magnitude is roughly one order of magnitude larger than $\sigma^{(2)}_{zzz}$ and several times larger than $\sigma^{(2)}_{yyz}$. This again indicates a strong coupling between in-plane optical excitation and out-of-plane charge redistribution, in line with the monolayer results but with a much stronger absolute response.

    The static field also renormalizes the electronic structure. The direct band gap decreases approximately linearly with the applied field, with a slope of about $-3.23~\mathrm{eV}/(\mathrm{V}/\text{\AA})$, in agreement with photoluminescence experiments and simulations \cite{liu_tuning_2012, chu_electrically_2015, sun_measuring_2023}. This redshift produces a clear drift of the optical peaks with increasing $E_{\mathrm{DC}}$. The most intense peaks in $\sigma^{(2)}$ appear well above the absorption edge, around $1.3$ to $1.5E_g$, in agreement with GW--BSE studies of the linear optical spectrum of \ce{MoS2} \cite{qiu_optical_2013}. Those works showed that the absorption in this range is dominated by higher-lying excitations with significant contributions from states away from the $K$ valleys and with orbital characters beyond the simple $d_{z^2}$ conduction-band edge. The independent-particle approximation does not capture exciton binding energies or any sub-gap excitations, but the enhanced response is consistent with the presence of such strongly allowed transitions.

    The detailed behavior at low and intermediate fields is shown in Fig. \ref{fig:linearity}.
    For weak static fields, $\sigma^{(2)}_{yyy}$ grows linearly with the static field up to $|E_{\mathrm{DC}}|\lesssim 1$ mV/\AA, whereas components that contain out-of-plane indices remain linear up to approximately $|E_{\mathrm{DC}}|\lesssim 7.5$ mV/\AA. For larger fields the response saturates. The in-plane components exhibit a sharper inflection, whereas the out-of-plane ones saturate more gradually. The saturation field also depends on photon energy, so different frequencies show different onset values.
    
    In this linear regime, the response satisfies $\sigma^{(2)}_{zyy}(\omega;E_{\mathrm{DC}})\approx \sigma^{(3)}_{zyyz}\, E_{\mathrm{DC}}$ as expected from the Taylor expansion discussed in Sec.~\ref{sec:theory}. The derivative with respect to $E_{\mathrm{DC}}$ is nearly constant in this range, which confirms the direct link between the induced second-order response and the third-order conductivity. At larger fields the derivative develops strong oscillations, indicating the breakdown of the first-order Taylor expansion and the growing importance of higher-order terms as hoppings and intersite dipoles depend nonlinearly on $E_{\mathrm{DC}}$.
    
\begin{figure}[b]
        \centering
        \includegraphics[width=\linewidth]{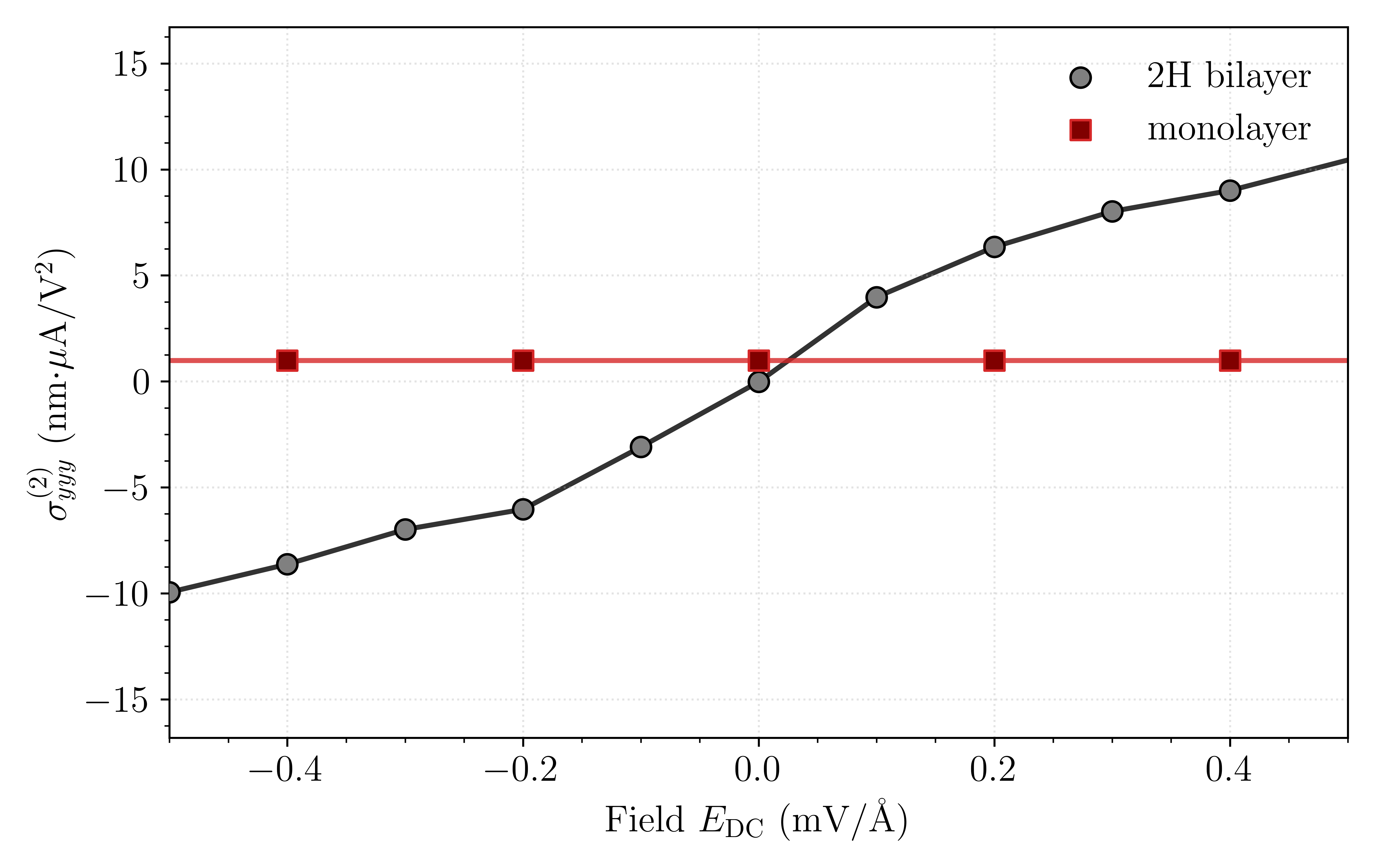}
        \caption{
        Comparison between the field-induced second-order conductivity of centrosymmetric 2H (AA$'$) bilayer \ce{MoS2} (black circles) and the intrinsic monolayer response (red squares) at $\hbar\omega\approx E_g$. The bilayer exhibits a strong (essentially) linear dependence on $E_{\mathrm{DC}}$, while the monolayer remains nearly constant. The sign reversal for opposite field polarities reflects the antisymmetric character of the field-induced shift current, which vanishes in the centrosymmetric limit.
        }
        \label{fig:comparison}
    \end{figure}

    It is instructive to compare the centrosymmetric 2H bilayer with the monolayer, as shown in Figure \ref{fig:comparison}. The monolayer response is almost insensitive to $E_{\mathrm{DC}}$, while in the bilayer the same static field activates a strong and tunable $\sigma^{(2)}$. An analogous distinction between monolayer and bilayer responses is seen in electric-field-controlled SHG experiments on bilayer \ce{MoS2} \cite{klein2017electric}, where the monolayer exhibits little variation with applied field while the bilayer displays a strong linear increase. In nonlinear optics it is common to write an effective susceptibility $\chi^{(2)}_{\mathrm{eff}}\propto \chi^{(3)} E_{\mathrm{DC}}$, which leads to an SHG intensity that scales as $I_{2\omega}\propto E_{\mathrm{DC}}^2$. In the present case an analogous mechanism appears in the photoconductivity. The field-dressed Hamiltonian produces an effective $\sigma^{(2)}_{\mathrm{eff}}\propto E_{\mathrm{DC}}$, which yields a dc current that is linear in the external static field and quadratic in the optical field. Field-induced shift current and EFISH therefore emerge from the same symmetry-breaking process \cite{chen_gigantic_2019, fan_electric-field-induced_2025}.
    
\subsection{3R-\ce{MoS2} Bilayer}

    The 3R-stacked (AB) bilayer \ce{MoS2} belongs to the non-centrosymmetric $C_{3v}$ point group, which is the same symmetry that the monolayer and the 2H bilayer acquire only after inversion is broken. As a result, the shift current is already finite at zero static field, with nonzero tensor components given by Eq. \ref{eq:tensor_components}. A perpendicular static field $E_{\mathrm{DC}}$ does not change the point group, but it modifies the potential difference between the layers and the degree of interlayer hybridization. In this case the field acts as a tuning parameter for the second-order response rather than as a symmetry-enabling perturbation.
    
\begin{figure}[ht]
        \centering
        \includegraphics[width=\linewidth]{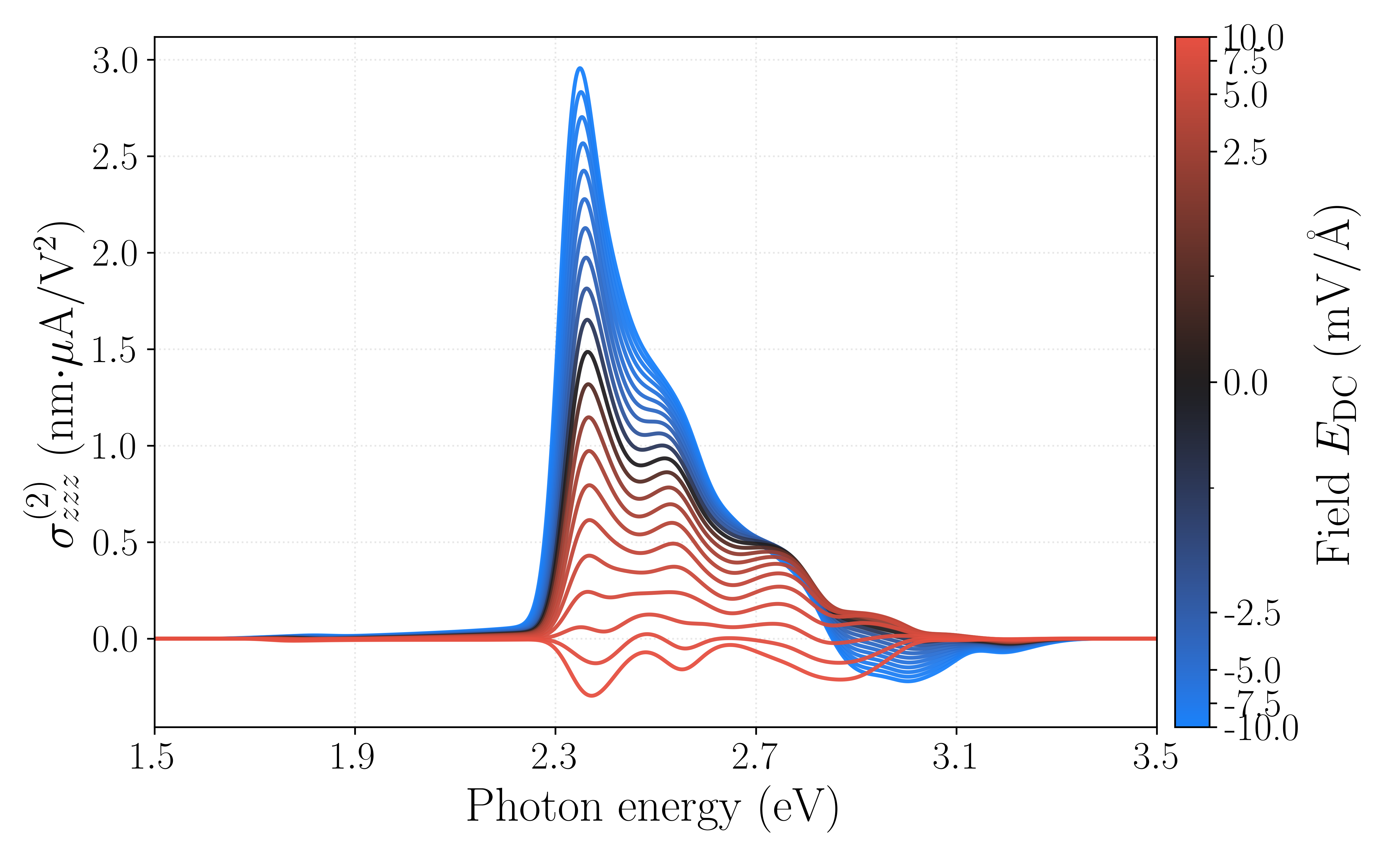}
        \caption{
        Field dependence of $\sigma^{(2)}_{zzz}(\omega;E_{\mathrm{DC}})$ in 3R-stacked \ce{MoS2}. 
        Positive and negative static fields modulate the intrinsic response in opposite directions, and a strong suppression appears when the external field compensates the built-in polarity of the stacking.
        }
        \label{fig:mos2_AB_zzz}
    \end{figure}

    For the in-plane response, the effect of $E_{\mathrm{DC}}$ is similar to the monolayer, since components such as $\sigma^{(2)}_{yyy}$ remain nearly unchanged as the field increases. The main difference appears in tensor elements that contain out-of-plane indices. These components are highly sensitive to the magnitude and sign of $E_{\mathrm{DC}}$ and display a clear antisymmetric dependence, which is consistent with the odd character of the perturbation $-eE_{\mathrm{DC}}\hat r_z$. A representative case is shown in Fig. \ref{fig:mos2_AB_zzz} for the $\sigma^{(2)}_{zzz}$ component.
    
    Unlike the 2H bilayer, where all components vanish in the centrosymmetric limit and grow from zero once inversion is broken, the 3R stacking already supports intrinsic shift conductivity in every symmetry-allowed component. When the external field is applied, the perturbation $-eE_{\mathrm{DC}}\hat r_z$ can either reinforce or oppose the built-in polarity of the stacking. One field orientation enhances the intrinsic asymmetry and increases the shift conductivity, while the opposite polarity partially compensates the internal potential difference. At a particular field strength the two contributions nearly cancel and the effective layer asymmetry is strongly reduced. In this situation the shift current is suppressed and the response mimics a temporary restoration of a mirror-like character across the $xy$ plane.
    
    This compensation effect has no analogue in the monolayer or in the 2H bilayer. In the monolayer, in-plane components are almost insensitive to $E_{\mathrm{DC}}$ and out-of-plane components only appear once the field breaks the horizontal mirror. In the 2H stacking, the static field activates a response that is strictly zero at $E_{\mathrm{DC}}=0$ and initially grows linearly with the field. The 3R structure therefore represents an intermediate case, where an external static field can either amplify or quench an already finite shift current. The $\sigma^{(2)}_{zzz}$ component, which does not couple to in-plane optical fields, provides a particularly clear example of this behavior and is shown in Fig. \ref{fig:mos2_AB_zzz}.

    Similar field–polarity compensation effects have been reported in ferroelectric photovoltaic devices and gate-tunable van der Waals heterostructures, where external fields or gate voltages can enhance, suppress, or even reverse photocurrents by competing with built-in fields \cite{li2021enhanced, Bai2024_CIPS_Reversal}. However, to the best of our knowledge, an analogous compensation mechanism has not been explicitly demonstrated for shift-current responses in non-ferroelectric 3R-stacked transition-metal dichalcogenides.
    
\section{Conclusions}

    We have shown that a static out-of-plane electric field provides an effective way to enable and tune shift-current responses in two-dimensional materials. Using a combined DFT-Wannier-XATU.Optix workflow \cite{XATU}, we demonstrated how the field breaks inversion in the 2H bilayer and activates a strong second-order response that grows linearly at small static fields and saturates at larger fields. The monolayer and 3R bilayer remain in the same $C_{3v}$ class under the field, but their responses differ markedly because the field either generates new out-of-plane components in the monolayer or tunes an already finite shift current in the polar 3R stacking.
    
    The onset and saturation of the field-induced response occur well below the dielectric breakdown limit. The field strengths explored here extend from $1$ mV/\AA to $10$ mV/\AA to illustrate the full evolution of the response, although the relevant behavior appears within experimentally accessible values below $0.1$ mV/\AA. These trends provide a microscopic basis for the field control of photogalvanic effects in devices and broaden the range of nonlinear optical phenomena available in centrosymmetric layered materials.

\section*{Acknowledgments}
    J.J. Esteve-Paredes thanks Daniel J. Passos for helpful discussions. This work has been funded by the Spanish Ministry of Science, Innovation and Universities \& the State Research Agency (MCIU/AEI/FEDER,UE) through grants TED2021-131323BI00 and PID2022-141712NB-C21, the "María de Maeztu" Programme for Units of Excellence in R\&D (CEX2023-001316-M),  the  Comunidad de Madrid within the Recovery, Transformation and Resilience Plan, and by NextGenerationEU programme from the European Union through the project "Disruptive 2D materials” (MAD2D-CM-UAM7) and the Generalitat Valenciana through the Program Prometeo (2021/017). We also acknowledge computer resources and assistance provided by Centro de Computación Científica de la Universidad Autónoma de Madrid and RES resources (FI-2025-1-12, FI-2024-3-0010, FI-2024-2-0016, FI-2024-1-0038, FI-2023-3-0049, and FI-2023-2-0013). 
    Guilherme I. Janone and Wendel S. Paz acknowledge financial support from the Brazilian funding agencies FAPES (1044/2022, 1081/2022 – P:2022-8L35F, and 875/2023 – P:2023-V36VC) and CNPq (under grants 444450/2024-6, 305227/2024-6, and 442781/2023-7). They are also grateful for the computational resources provided by the Sci-Com Lab/UFES. Guilherme J. Inacio expresses his gratitude to Darian Leucian Pele for fruitful discussions.

\begin{appendices}
\section{Computational Methods}\label{sec:methods}

\subsection{Electronic structure}
    
    The electronic structure of \ce{MoS2} was obtained from density functional theory calculations performed with the \textsc{Quantum Espresso} package \cite{giannozzi2009quantum}. 
    Exchange-correlation effects were described within the generalized gradient approximation using the Perdew-Burke-Ernzerhof functional \cite{perdew1996generalized}. 
    Core-valence interactions were treated with KJPAW pseudopotentials. 
    A plane-wave cutoff of $50\,\mathrm{Ry}$ for the wavefunctions and $310\,\mathrm{Ry}$ for the charge density was used together with an $18\times 18\times 1$ Monkhorst-Pack grid. 
    Atomic positions were relaxed until residual forces were smaller than $10^{-3}\,\mathrm{(a.u)}$ and the change in total energy was smaller than $10^{-4}\,\mathrm{(a.u)}$. A vacuum spacing of $20$ \AA\ was included to avoid spurious interactions between periodic replicas, and 
    van-der-Waals dispersion forces were included through D3 Grimme corrections \cite{grimme2006semiempirical}.
    
    \subsection{Wannier Hamiltonian}
    
    The low-energy Hamiltonian was constructed using the \textsc{Wannier90} code \cite{mostofi2014updated}. 
    Closest Wannier Functions (CWFs) \cite{ozaki2024closest,oiwa_symmetry-adapted_2025} were employed in place of maximally localized Wannier functions to preserve atomic character and ensure gauge continuity in the presence of the static field. 
    Mo $s$, $p$ and $d$ orbitals, and S $s$ and $p$ orbitals were used as projectors, and the disentanglement window was selected with a Fermi-Dirac smearing ranging from $T_0=10^{-6}$ K to $T_1=5$ K. 
    No maximal localization procedure was applied. 
    The resulting Hamiltonian reproduces the DFT band structure near the Fermi level and provides well-behaved intersite position-matrix elements $r^{(z)}_{\alpha\alpha'\mathbf{k}}$ needed for the static field coupling.
    
    \subsection{Optical and shift-conductivity calculations}
    
    The linear and nonlinear conductivities were obtained within the independent-particle approximation using the open-source \textsc{XATU.Optix} package \cite{XATU, esteve-paredes_excitons_2025, uria2023efficient}. 
    \textsc{XATU.Optix} evaluates the second-order tensor $\sigma^{(2)}_{abc}(0;\omega,-\omega)$ from the velocity and dipole matrix elements of the Wannier Hamiltonian. 
    All spectra were computed using a Lorentzian broadening of $\hbar\eta=0.030\,\mathrm{eV}$ and sampling over $161$ neighbouring cells in the Brillouin zone. 
    Excitonic effects were not included so that the influence of the static field on the single-particle band structure can be isolated.
\end{appendices}

\bibliographystyle{ieeetr}
\bibliography{biblio}

\end{document}